\documentclass[journal]{IEEEtran}

\usepackage{cite}
\usepackage{amsmath,amssymb,amsfonts}
\usepackage{graphicx}
\usepackage{booktabs}
\usepackage{multirow}
\usepackage{array}
\usepackage{xcolor}
\usepackage{tikz}
\usetikzlibrary{arrows.meta,positioning,fit,calc}
\usepackage[hidelinks]{hyperref}

\DeclareUnicodeCharacter{202F}{\,}
\begin{document}

\title{Evaluation of portability and performance of an OpenMP5 offloaded Quantum-Inspired Evolutionary Optimization Across the GPU Ecosystem}

\author{Kasturi~Venkata~Srikanth$^{*}$,
        Ashish Singh$^{*}$,
        Ferdin~Sagai~Don~Bosco$^{*}$,
        Aman~Mittal,
        Abhishek~Singh,
        Aditya~Singh
        and~Abhishek Chopra
\thanks{$^{*}$These authors contributed equally to this research.}%
\thanks{All authors are with BosonQ Psi Corporation, Bangalore, Karnataka, India.}%
\thanks{K.~V. Srikanth (ORCID: 0000-0002-0159-773X, kasturi.srikanth@bqpsim.com);
Ashish Singh (ORCID: 0009-0005-9365-2837, ashishsinghrajput107@gmail.com);
F.~S.~D. Bosco (ORCID: 0000-0002-5507-5304, ferdindon@bqpsim.com);
A.~Mittal (ORCID: 0009-0002-7737-6893, aman.mittal@bqpsim.com);
Abhishek Singh (ORCID: 0009-0009-7140-7047, abhishek.singh@bqpsim.com);
Aditya Singh (ORCID: 1234-5678-9012, singh.aditya@bqpsim.com);
A.~Chopra (ORCID: 0000-0002-1196-7765, abhishekchopra@bqpsim.com).}}

\markboth{}%
{Srikanth \MakeLowercase{\textit{et al.}}: Portable QIEO on the GPU Ecosystem}

\maketitle

\begin{abstract}
Quantum-inspired evolutionary optimization (QIEO) is a new class of population-based metaheuristic optimization algorithms which represents design variables as a set of qubits and searches a continuous, multi-dimensional landscape through rotation of the qubit's amplitude pair. Every generation rotates those amplitudes toward a single elite, which corresponds to that generation's best. The per-generation cost scales as $O(N_p N_g)$ for $N_p$ chromosomes and $N_g$ genes (decision variables). 

The key kernels of QIEO, namely, Quantum Information Processing (QIP), Function Evaluation (FE), DetermineElite, and Classical Information Processing (CIP) are investigated in chromosome-parallel offload, in which one GPU thread is assigned to one complete chromosome, and gene-parallel offload, where multiple cooperating threads are assigned to the genes within a single chromosome. The former exposes candidate solutions to parallelism, whereas the latter additionally exposes work within each solution to parallelism.

Production use of such solvers is rarely confined to a single machine class. Prototypes are run on laboratory servers- before moving to rented cloud workstations for more involved campaigns. The largest problems are reserved for leadership-class accelerators. This paper asks whether a \emph{single} OpenMP~5 source of QIEO, offloaded with \texttt{\#pragma omp target}, is a viable production path in each of those settings.

We report three independent, campaigns of the 0/1 knapsack problem against a same-source multi-core Intel CPU baseline. The study comprises approximately 3,000 runs spanning varying chromosome and gene counts, evaluated using both chromosome-level and gene-level offload strategies on the NVIDIA Tesla V100 SXM2, NVIDIA A100 80GB, and AMD Instinct MI300X GPUs. Deployment-specific nuances such as Volta's constant-memory cliffs, Ampere's L2 persistence and \texttt{cp.async}, CDNA~3's Infinity Cache and XCD occupancy are addressed to ensure high performance of these platforms.

Results reveal gene-parallel offload achieved geometric-mean speedups of 90$\times$, 136$\times$, and 155$\times$ over a single CPU core on the V100, A100, and MI300X, respectively, and 12$\times$, 17$\times$, and 16.6$\times$ over 72 host threads. In every campaign, chromosome-parallel offload is slower in 100\% of matched pairs. Furthermore DetermineElite, the $O(N_p)$ selection of the generation-best chromosome, is found to be better suited to the host than to the device.
\end{abstract}


\section{Introduction}

Combinatorial optimization on classical hardware remains the workhorse of logistics, packing, and resource-allocation pipelines even as quantum-inspired heuristics grow in popularity~\cite{han2002qea,eswara2024qieo}. In a quantum-inspired evolutionary optimizer (QIEO) , each gene is a qubit amplitude pair $(\alpha,\beta)$ with $|\alpha|^2+|\beta|^2=1$, which collapses to a binary allele upon observation. A rotation gate then steers amplitudes toward elite solutions~\cite{han2002qea}. The resulting method is a generational loop whose arithmetic intensity is modest but whose parallelism is abundant, which is
only useful if it can be realized where practitioners actually run the solver. In industrial and academic practice the same code must survive three quite different ecosystems:

\begin{enumerate}
\item an \emph{in-house server}, typically a previous-generation accelerator already installed in a laboratory rack, used for development, regression, and modest campaigns;
\item a \emph{cloud workstation}, rented by the hour, with a contemporary high-memory GPU and a software stack that the user does not own;
\item a \emph{leadership-class} node, allocated through a center or a large-scale cloud reservation, whose memory capacity and compute-unit count so large that the largest $(N_p,N_g)$ instances too remain resident.
\end{enumerate}

CUDA studies of QIEO have already shown that memory placement and thread-block geometry dominate runtime on NVIDIA devices~\cite{mittal2025qieo}. Those results do not answer the production questions that this paper treats as primary:

\textbf{Can the algorithm expressed portably in OpenMP~5 target offload~\cite{openmp50}, deliver a decisive advantage over a same-source multi-core CPU in each of the three environments above?}
 
The thesis is therefore one of \emph{deployment adequacy}, not of inter-GPU ranking. 

A portable OpenMP~5 formulation of QIEO kernels is developed and fine-tuned for the Tesla V100~SXM2 (in-house), A100~80GB (cloud workstation), and Instinct MI300X (leadership class). Campaigns of 0/1 Knapsack problem spanning different chromosome and gene counts are performed. Each run is also subject to two offload strategies; chromosome-parallel and gene-parallel, totalling 2940 measured runs.

The contributions are:
\begin{enumerate}
\item Portable version of product that is GPU agnostic and capable of running on varied ecosystems (for free trial contact please reach out to bizdev@bqpsim.com)

\item A synthesis of architecture, performance-in-role, and workflow placement for the three GPUs (Section~\ref{sec:three}), without an inter-GPU ranking.

\item A deployment playbook that translates those portable rules onto Volta's 64\,KiB constant memory, Ampere's 164\,KiB shared memory, 40\,MB L2 persistence window and \texttt{cp.async} pipeline, and CDNA~3's 304 compute units, 192\,GB HBM3 and 256\,MB Infinity Cache.
\end{enumerate}

0/1 Knapsack remains the combinatorial probe~\cite{garey1979,jooken2022,zhou2023} used in this study.
The results demonstrate the versatility of an OMP5 offloaded QIEO across the diversified GPU ecosystem. Ensuring portability while retaining performance can be achieved by necessitating gene-parallel offload in every tested ecosystem. Moreover, the computationally expensive fitness evaluation process is strongly dependent on data persistence on the on-chip memory, managing which is essential for high performance. Lastly, DetermineElite should remain a host-side reduction after Evaluation and before CIP, as it performs poorly when offloaded to a GPU.

\section{Related Work}

\noindent\textbf{Takeaway.}
\textbf{Earlier GPU QIEO papers mostly used CUDA on one NVIDIA card.
This paper keeps one OpenMP source and asks whether that source is good enough in three real places: lab, cloud, and a leadership node.}

Han and Kim introduced quantum-inspired evolutionary algorithms for combinatorial problems, including knapsack, using qubit chromosomes and rotation gates~\cite{han2002qea}. Eswara Sai \emph{et al.} formulated the QIEO variant whose kernel split (QIP, FE, CIP) we inherit~\cite{eswara2024qieo}. GPU ports have largely been CUDA-centric~\cite{mittal2025qieo,nowotniak2014gpu,patil2024bench}. 

However, heterogeneous nature of modern HPC ecosystems demand portability in developed solvers without compromising performance. OpenMP~5 target offload is the source-level contract~\cite{openmp50} that has shown great promise in providing the necessary versatility. The present research aims to test QIEO on the Volta~\cite{volta2018,v100ds}, Ampere~\cite{a100wp,a100ds,amdmi300x}, and CDNA~3~\cite{rocmomp} architectures.

The present paper is complementary to the single-GPU studies. It holds the algorithm and the programming model fixed, changes only the \emph{ecosystem} in which the binary runs. It strongly refuses the inter-GPU ranking that those ecosystems would otherwise invite and applies a "right tool for the right job" mindset.

\section{Problem statement}

\subsection{Formulation}
Knapsack remains a standard NP-complete stress test for metaheuristics~\cite{garey1979,jooken2022,zhou2023}; here it is used because its arithmetic is transparent and its data motion is exactly the $N_p\times N_g$ grid that QIEO must stream every generation.

A chromosome $c=1,\ldots,N_p$ stores $N_g$ qubits. Observation at generation $t$ produces a binary vector $x^{(c)}\in\{0,1\}^{N_g}$. For knapsack item $i$ with value $v_i$ and weight $w_i$, the fitness of chromosome $c$ is
\begin{equation}
f(x^{(c)})
=
\begin{cases}
\displaystyle\sum_{i=1}^{N_g} v_i x^{(c)}_i,
& \text{if }\displaystyle\sum_{i=1}^{N_g} w_i x^{(c)}_i \le W,\\[6pt]
-\lambda\Bigl(\displaystyle\sum_{i=1}^{N_g} w_i x^{(c)}_i - W\Bigr),
& \text{otherwise,}
\end{cases}
\label{eq:fitness}
\end{equation}
where $W$ is the knapsack capacity and $\lambda>0$ is a penalty for weight constraint violation. Equation~\eqref{eq:fitness} is the Evaluation kernel consisting of $O(N_g)$ independent multiply-adds per chromosome, followed by a compare against $W$. 

\subsection{Methodology}
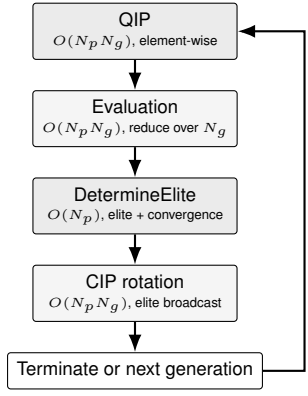
\begin{figure}[!t]
\centering
\begin{tikzpicture}[
  font=\scriptsize\sffamily,
  box/.style={draw, rounded corners=1.2pt, align=center, inner sep=3.5pt, minimum width=2.7cm},
  arr/.style={-{Latex}, thick}
]
\node[box, fill=black!8] (qip) {QIP\\[-1pt]{\tiny $O(N_p N_g)$, element-wise}};
\node[box, fill=black!4, below=4.0mm of qip] (fe) {Evaluation\\[-1pt]{\tiny $O(N_p N_g)$, reduce over $N_g$}};
\node[box, fill=black!8, below=4.0mm of fe] (de) {DetermineElite\\[-1pt]{\tiny $O(N_p)$, elite + convergence}};
\node[box, fill=black!4, below=4.0mm of de] (cip) {CIP rotation\\[-1pt]{\tiny $O(N_p N_g)$, elite broadcast}};
\node[box, fill=white, below=4.0mm of cip] (stop) {Terminate or next generation};
\draw[arr] (qip) -- (fe);
\draw[arr] (fe) -- (de);
\draw[arr] (de) -- (cip);
\draw[arr] (cip) -- (stop);
\draw[arr] (stop.east) -- ++(0.55,0) |- (qip.east);
\end{tikzpicture}
\caption{Instrumented QIEO generation. QIP and CIP are dense element-wise updates of the qubit grid; Evaluation reduces along genes; DetermineElite reduces along chromosomes and precedes CIP.}
\label{fig:kernels}
\end{figure}

The production code times four regions that together reconstruct the per-generation wall time (Fig.~\ref{fig:kernels}):

\begin{itemize}
\item \textbf{Quantum Information Processing (QIP)}. Resample qubit amplitudes and collapse them to binary alleles. Arithmetic is element-wise over the $N_p\times N_g$ qubit array: $O(N_p N_g)$ with unit-stride stores if the layout is consistent with the parallel axis.
\item \textbf{Evaluation} (fitness evaluation). Implements eq.~\eqref{eq:fitness}. Complexity $O(N_p N_g)$ but with a \emph{reduction along the gene axis} for each chromosome.
\item \textbf{DetermineElite}. A population-wide $O(N_p)$ reduction that selects the generation-best chromosome (elite index and string) after Evaluation, so that CIP can rotate toward that elite. The same scalar feeds the convergence test (fitness plateau over a window of generations). The operation is two to four orders of magnitude less work than the dense kernels.
\item \textbf{Classical Information Processing (CIP)}. Apply a lookup-table rotation to each qubit toward the elite already chosen by DetermineElite: a dense $O(N_p N_g)$ pass with a broadcast of the elite string.
\end{itemize}

The measured identity
\begin{equation}
T_{\mathrm{gen}}
\;\approx\;
T_{\mathrm{QIP}}+T_{\mathrm{FE}}+T_{\mathrm{DE}}+T_{\mathrm{CIP}}+T_{\mathrm{launch}}
\label{eq:budget}
\end{equation}
holds to within a few percent. $T_{\mathrm{DE}}$ is DetermineElite. $T_{\mathrm{launch}}$ captures target enter/exit, implicit barriers, and host-side bookkeeping. It is negligible on large grids and decisive on tiny ones.

\subsection{OMP5 Offload Strategies}
Let $i$ index chromosomes and $j$ index genes. The same OpenMP~5 source~\cite{openmp50} is compiled for the host and for every device in this study. Host runs set the thread count $p$ (\texttt{Parallel\_$p$} in the archives). Device runs use \texttt{\#pragma omp target}. A command-line switch selects one of two loop mappings; the binaries are prefixed \texttt{Chromosome\_} and \texttt{Gene\_}.

Chromosome-parallel offload parallelizes only $i$,
\begin{equation}
\texttt{parallel for } i=1\ldots N_p
\quad\text{with a serial inner loop over }j,
\label{eq:chrommap}
\end{equation}
so at most $N_p$ workers are busy and each still walks $O(N_g)$ genes. Gene-parallel offload adds \texttt{collapse(2)} over $(i,j)$, exposing up to $N_p N_g$ independent updates for QIP and CIP; Evaluation then needs a segmented reduction along $j$. On a CPU the two mappings differ mainly in cache reuse once $N_p$ oversubscribes the socket. On a GPU the distinction is occupancy: $N_p$ waves cannot fill 80--304 compute units unless the population is huge. Later sections measure that gap; they do not change the source.

A gene-parallel QIP has the shape
\begin{verbatim}
#pragma omp target teams distribute \
    parallel for collapse(2) \
    map(to: v[0:Ng], w[0:Ng]) \
    map(tofrom: Q[0:Np*Ng], x[0:Np*Ng])
for (int i = 0; i < Np; ++i)
  for (int j = 0; j < Ng; ++j)
    reset_qubit(Q, x, i, j);
\end{verbatim}
Weights, values, and $Q$ stay mapped across generations, so after the first HtoD the interconnect is no longer the critical path. The four timed regions in~\eqref{eq:budget} still insert an implicit barrier at each target boundary. The portable fix is structural, not vendor-specific: keep $Q$ on the device, run DetermineElite on the host after Evaluation and before CIP, and treat constant-memory, L2-window, and \texttt{cp.async} helpers as optional last-factor steps on a given GPU.

\subsection{Deployment Environments}

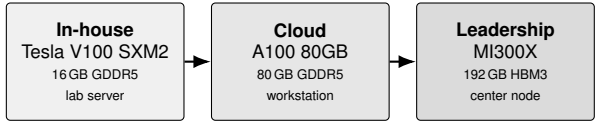
\begin{figure}[!t]
\centering
\begin{tikzpicture}[
  font=\scriptsize\sffamily,
  box/.style={draw, rounded corners=1.5pt, align=center, inner sep=4pt,
              minimum width=2.35cm, minimum height=1.55cm},
  arr/.style={-{Latex}, thick}
]
\node[box, fill=black!6] (v) {\textbf{In-house}\\Tesla V100 SXM2\\{\tiny 16\,GB GDDR5}\\{\tiny lab server}};
\node[box, fill=black!10, right=3.5mm of v] (a) {\textbf{Cloud}\\A100 80GB\\{\tiny 80\,GB GDDR5}\\{\tiny workstation}};
\node[box, fill=black!14, right=3.5mm of a] (m) {\textbf{Leadership}\\MI300X\\{\tiny 192\,GB HBM3}\\{\tiny center node}};
\draw[arr] (v) -- (a);
\draw[arr] (a) -- (m);
\end{tikzpicture}
\caption{Deployment ecosystem used in this study. The same OpenMP~5 QIEO source is evaluated in each environment against that environment's CPU baseline. Arrows denote a typical workflow, not a performance ranking.}
\label{fig:ecosystem}
\end{figure}

Figure~\ref{fig:ecosystem} situates the machines. Table~\ref{tab:hw} lists the constraints that a kernel must respect in each environment. The table is a \emph{budget}, not a ranking. 

\begin{table*}[!t]
\centering
\caption{Deployment environments. Each device campaign is scored only against the Intel OpenMP host that produced that archive.}
\label{tab:hw}
\renewcommand{\arraystretch}{1.15}
\begin{tabular}{>{\raggedright\arraybackslash}p{0.22\textwidth}
                >{\raggedright\arraybackslash}p{0.24\textwidth}
                >{\raggedright\arraybackslash}p{0.24\textwidth}
                >{\raggedright\arraybackslash}p{0.22\textwidth}}
\toprule
 & \textbf{In-house server} & \textbf{Cloud workstation} & \textbf{Leadership node} \\
\midrule
Accelerator & NVIDIA Tesla V100 SXM2 & NVIDIA A100 80GB & AMD Instinct MI300X \\
Architecture & Volta GV100, CC 7.0 & Ampere GA100, CC 8.0 & CDNA~3, 8 XCDs \\
Parallel grain & 80 SMs, warp 32 & 108 SMs, warp 32 & 304 CUs, wavefront 64 \\
Device memory & 16\,GB GDDR5, $\approx 900$\,GB/s & 80\,GB GDDR5, $\approx 2.0$\,TB/s & 192\,GB HBM3, 5.3\,TB/s \\
On-chip reuse & 64\,KiB constant; $\approx 48$\,KiB shared typical & 64\,KiB constant; 164\,KiB shared; 40\,MB L2 & 256\,MB Infinity Cache \\
Host baseline & Multi-core Intel, OpenMP $p\le 72$ & Multi-core Intel, OpenMP $p\le 72$ & Multi-core Intel, OpenMP $p\le 72$ \\
Role in the story & Already-owned lab GPU & Rented high-memory workstation & Center-scale campaign node \\
\bottomrule
\end{tabular}
\end{table*}

The definition of a "Large" case varies as the GPU changes in Fig.~\ref{fig:ecosystem}. 

\begin{enumerate}
\item On V100 the $N_p\times N_g$ qubit grid is a \emph{capacity} problem: several Large cells in the archive have no timing footer because 16\,GB is exhausted. 
\item On A100~80GB the same grid is a \emph{locality} problem: the bytes fit, and the question is whether they sit in constant memory, a shared-memory tile, an L2 persistence window, or raw GDDR5. 
\item On MI300X the grid is an \emph{occupancy and Infinity-Cache} problem: 192\,GB admits instances that the other two environments cannot store, but 304 compute units still starve unless the gene axis is exposed.
\end{enumerate}

Read-only item tables occupy $8N_g$ bytes (two 32-bit integers). Using an 80\% occupancy safety margin~\cite{mittal2025qieo},
\begin{align}
N_{\mathrm{const}} &= \bigl\lfloor 0.8\cdot 64\cdot 1024 / 8\bigr\rfloor \approx 6553, \label{eq:nconst}\\
N_{\mathrm{sh,V100}} &= \bigl\lfloor 0.8\cdot 48\cdot 1024 / 8\bigr\rfloor \approx 4915, \label{eq:nshv}\\
N_{\mathrm{sh,A100}} &= \bigl\lfloor 0.8\cdot 164\cdot 1024 / 8\bigr\rfloor \approx 16793. \label{eq:nsha}
\end{align}
Equation~\eqref{eq:nconst} is shared by both NVIDIA environments. Equations~\eqref{eq:nshv}--\eqref{eq:nsha} are why a tile that evicts on Volta is legal on Ampere. On MI300X the analogous reuse structure is the 256\,MB Infinity Cache: even $N_g=200000$ ($1.6$\,MB of item tables) is a cache-resident object, while the qubit grid is not.

\section{Experimental Design}

\subsection{Workload}
\begin{figure}[!t]
\centering
\includegraphics[width=\columnwidth]{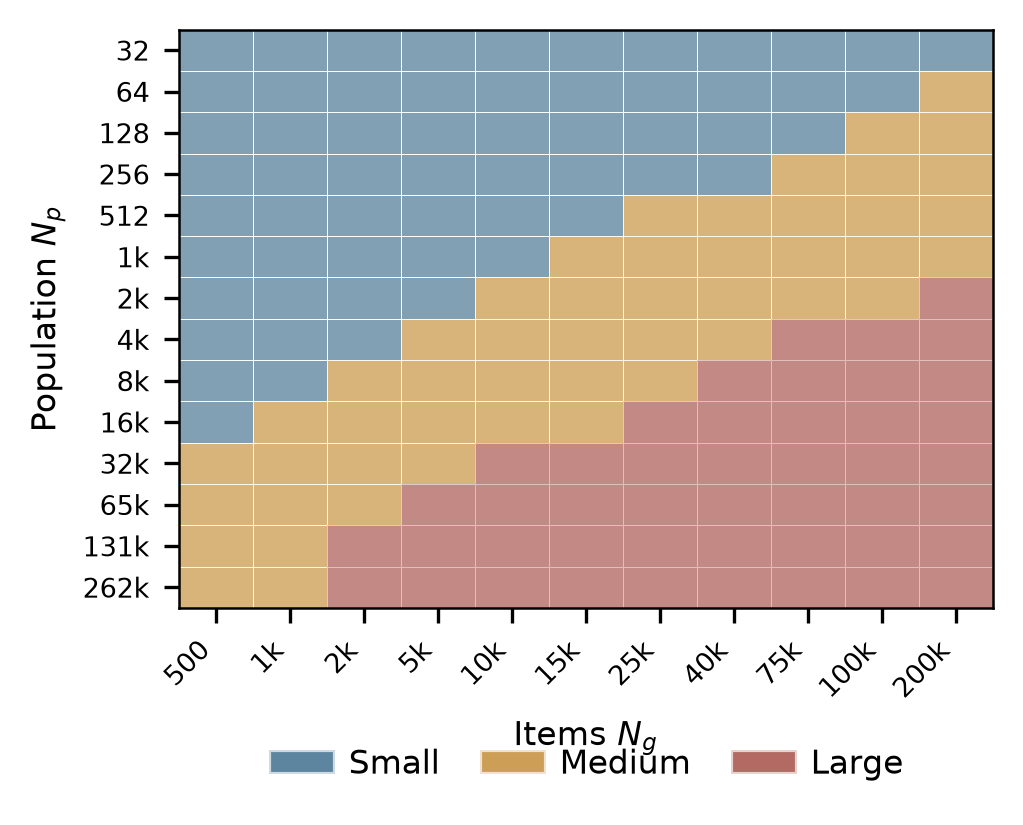}
\caption{Filename classes on the same $(N_p,N_g)$ grid as Fig.~\ref{fig:tgen_cpu}. Overlap is intentional: a cell may belong to more than one class. Archive pairs with no timing footer sit in the capacity corner and are coloured Large.}
\label{fig:classes}
\end{figure}

Instances are 0/1 knapsack problems. Each run is a point $(N_p,N_g)$: $N_p$ chromosomes (candidate packs) and $N_g$ items (genes). Tokens Small, Medium, and Large densify different corners of that plane (Fig.~\ref{fig:classes}). Small sits at modest populations, where occupancy and launch tax dominate and where the only full host thread sweep exists. Medium extends toward larger $N_p$ at mid-grid work. Large occupies the high-$N_p$, high-$N_g$ corner, where the qubit grid stresses device capacity and Evaluation owns $T_{\mathrm{gen}}$. Table~\ref{tab:classes} lists the unique-pair counts and ranges.

Convergence is declared when the elite fitness is stable for 20 generations and all fitnesses in that window lie within $10^{-6}$ of their mean; otherwise a generation cap (typically $10^4$) stops the run. Per-generation time $T_{\mathrm{gen}}$ is the figure of merit for hardware comparison because it is independent of the stochastic stopping time. Generations-to-stop is reported as a property of the instance, not of the device.

\begin{table*}[!t]
\centering
\caption{Filename size classes. Counts are unique $(N_p,N_g)$ pairs in the trial-1 archive. Classes densify corners of the plane; they are not a disjoint partition.}
\label{tab:classes}
\renewcommand{\arraystretch}{1.12}
\begin{tabular}{lccc p{0.42\textwidth}}
\toprule
\textbf{Class} & \textbf{Unique pairs} & \textbf{Population $N_p$} & \textbf{Items $N_g$} & \textbf{Role} \\
\midrule
Small & 59 & $32$--$16\,384$ & $500$--$200\,000$ & Occupancy, launch tax, and the only host sweep $p=1,2,4,\ldots,72$ \\
Medium & 49 & $64$--$262\,144$ & $500$--$200\,000$ & Mid-grid fill; larger populations. CPU points are $p=1$ only \\
Large & 46 & $2\,048$--$262\,144$ & $2\,500$--$200\,000$ & Capacity ceiling and Evaluation-dominated grids. CPU points are $p=1$ only \\
\bottomrule
\end{tabular}
\end{table*}

\subsection{Corpus}
Each campaign retains only the device files of that environment plus the shared CPU logs. Empty logs (no timing footer) are dropped rather than imputed. The retained counts are:
\begin{itemize}
\item in-house V100: \textbf{928} measured runs (736 CPU, 133 gene-parallel, 59 chromosome-parallel);
\item cloud A100: \textbf{990} runs (736 CPU, 146 gene-parallel, 108 chromosome-parallel);
\item leadership MI300X: \textbf{1022} runs (736 CPU, 150 gene-parallel, 136 chromosome-parallel).
\end{itemize}
Population sizes and item counts are
\begin{align*}
N_p &\in \{32,64,128,\ldots,262144\},\\
N_g &\in \{500,1000,2500,\ldots,200000\}.
\end{align*}
The full host thread sweep $p\in\{1,2,4,8,16,24,32,40,48,64,72\}$ exists for the Small class (59 matched pairs). Medium and Large CPU points are predominantly one-thread baselines, which is the correct reference for accelerator speedup.

\subsection{Metrics}
Let $T(p)$ be per-generation time on $p$ host threads of that campaign's CPU, and let $T_{\mathrm{G}}$, $T_{\mathrm{C}}$ be the gene- and chromosome-parallel device times. We report
\begin{align}
S_{\mathrm{G}} &= T(1)/T_{\mathrm{G}}, &
S_{\mathrm{C}} &= T(1)/T_{\mathrm{C}}, \label{eq:sg}\\
S_{72} &= T(72)/T_{\mathrm{G}}, &
\eta(p) &= \frac{T(1)}{p\,T(p)}, \label{eq:eff}\\
\rho &= T_{\mathrm{C}}/T_{\mathrm{G}}. \label{eq:rho}
\end{align}
Kernel fractions $T_k/T_{\mathrm{gen}}$ diagnose bottleneck shifts. Geometric means are used for speedups because the dynamic range spans three orders of magnitude. Rankings inside a campaign use $(N_p,N_g)$-matched triples. No speedup in this paper is a ratio of one GPU's $T_{\mathrm{gen}}$ to another's.

\section{Results}

The host baseline is reported first.
Device campaigns are then read by concern---end-to-end adequacy, occupancy, memory management, and the kernel-resolved picture---with per-GPU bullets and one learning paragraph per concern.

\subsection{Intel Xeon Platinum 8581C with OpenMP as a Shared Baseline}

The Intel Xeon Platinum 8581C is a high-core-count, 5th Generation "Emerald Rapids" server processor built on the Intel 10nm (Intel 7) process, primarily deployed in cloud environments like Google Compute Engine. Physically, the full chip has 60 cores and 120 threads, a base clock of 2.1 GHz, boosting up to 2.9-4.0 GHz, and a massive 300 MB L3 cache, but cloud providers typically expose it as a virtual CPU running at a fixed 2.30 GHz base frequency.

The Small-class thread sweep is common to the three archives and is the honest description of the CPU baseline.

\begin{figure*}[!t]
\centering
\includegraphics[width=0.98\textwidth]{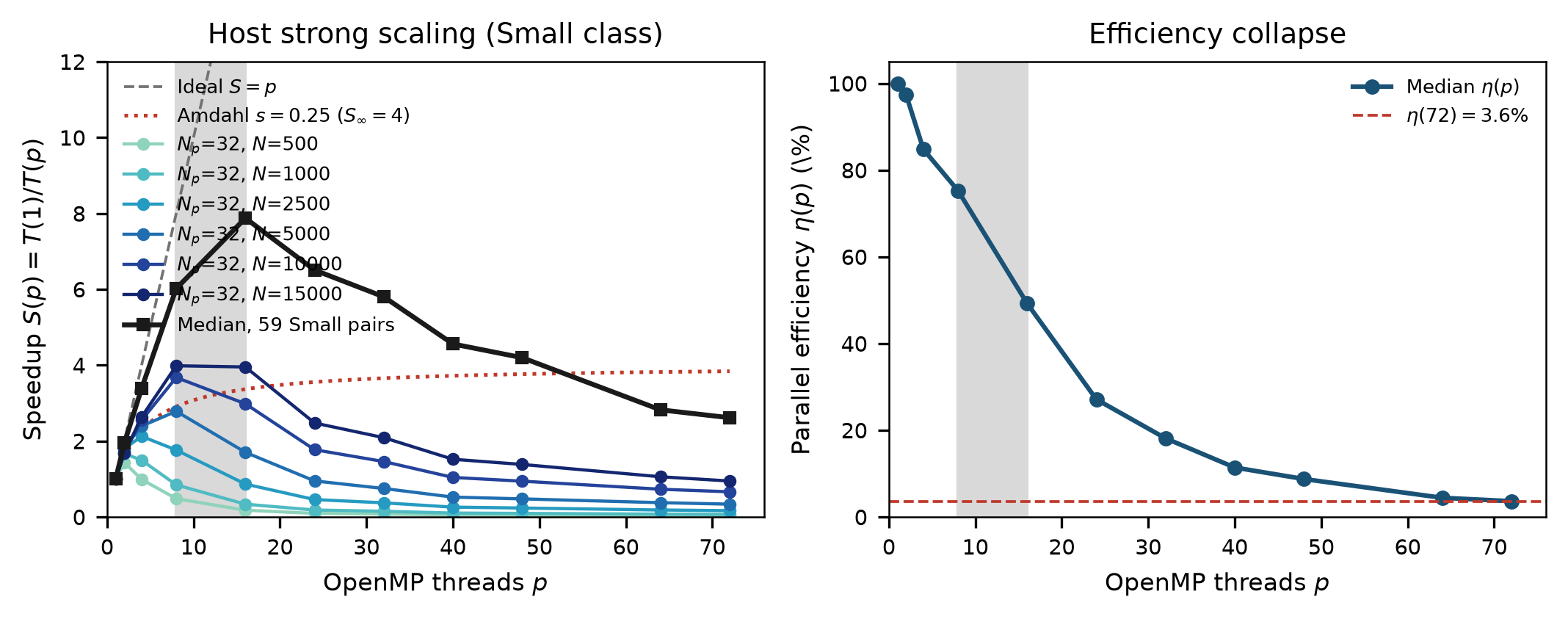}
\caption{Host OpenMP strong scaling on the Small class. Left: measured $S(p)$ for $N_p=32$ and the median over 59 matched pairs, against ideal $S=p$ and an Amdahl curve with $s=0.25$. The grey band is $p\in[8,16]$. Right: median parallel efficiency $\eta(p)=T(1)/(p\,T(p))$; the dashed line is $\eta(72)=3.6\%$.}
\label{fig:cpus}
\end{figure*}

Figure~\ref{fig:cpus} plots $S(p)=T(1)/T(p)$ against the OpenMP thread count. Speedup peels away from the ideal $S=p$ line immediately. For the densest family ($N_p=32$, $N_g\le 15000$) the curves hug an Amdahl envelope with serial fraction $s\approx 0.25$ ($S_\infty=4$)~\cite{amdahl1967,gustafson1988}. Across all 59 Small pairs the median peaks near $8\times$ at $p=16$, then recedes. At 72 threads the median speed-up is $2.62\times$, i.e.\ a parallel efficiency $\eta(72)=3.6\%$. The shaded band at $p\in[8,16]$ is therefore the production CPU mapping, not $p=72$.

This is Amdahl's law with a memory-bound serial fraction, plus a post-saturation collapse that pure Amdahl does not predict. QIP already consumes 77\% of serial $T_{\mathrm{gen}}$ (Table~\ref{tab:fracall}). Additional cores share the same DRAM controllers and saturate them. 

False sharing on the elite chromosome, OpenMP barriers between the four kernels, and NUMA placement of the $N_p\times N_g$ buffer compound the effect and drive $S(p)$ back toward unity past 16 threads. 

Larger $N_g$ helps slightly (more bytes per barrier) but does not restore efficiency, because the extra work is itself bandwidth-limited.

Two consequences follow for every later section. First, a 72-thread CPU is a weak opponent, which is why both $S_{\mathrm{G}}$ and $S_{72}$ are reported. Second, eight to sixteen host threads are the production CPU mapping; quoting only the one-core number would overstate the accelerator, and quoting only the 72-thread number would understate the host.

\begin{figure}[!t]
\centering
\includegraphics[width=\columnwidth]{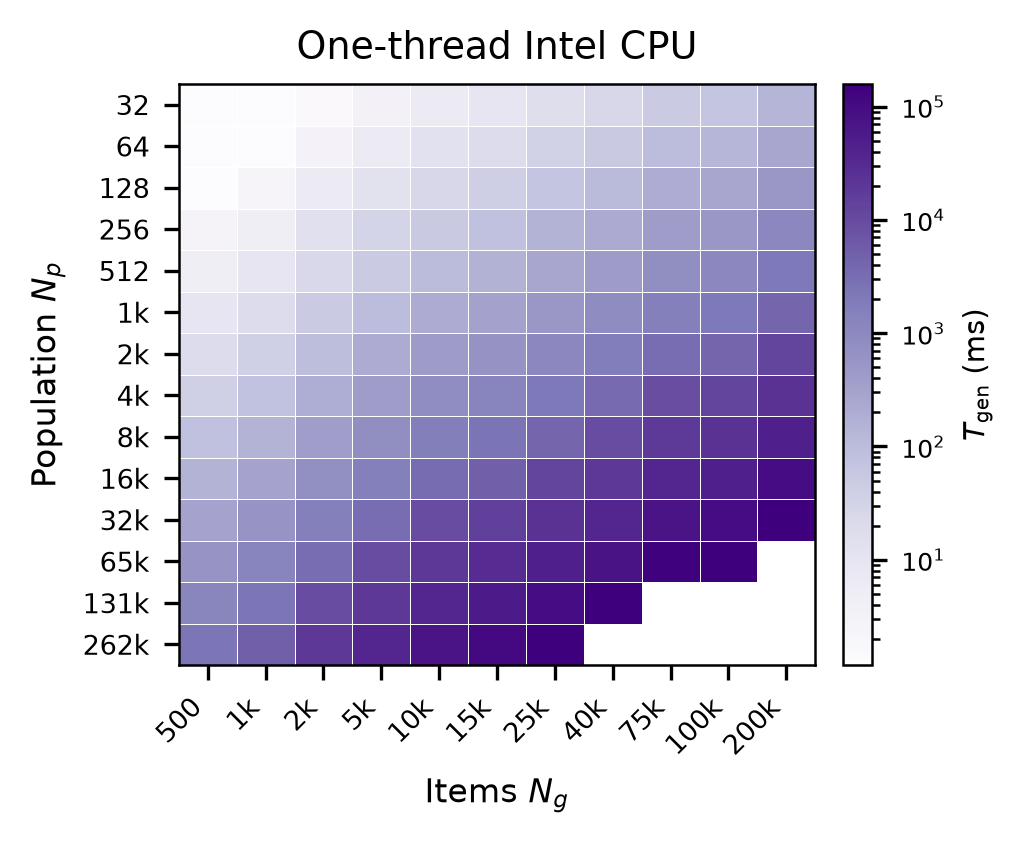}
\caption{Per-generation time on one Intel OpenMP thread. Colour is logarithmic. This is the $T(1)$ field used for $S_{\mathrm{G}}$ in every campaign.}
\label{fig:tgen_cpu}
\end{figure}

\begin{figure*}[!t]
\centering
\includegraphics[width=0.98\textwidth]{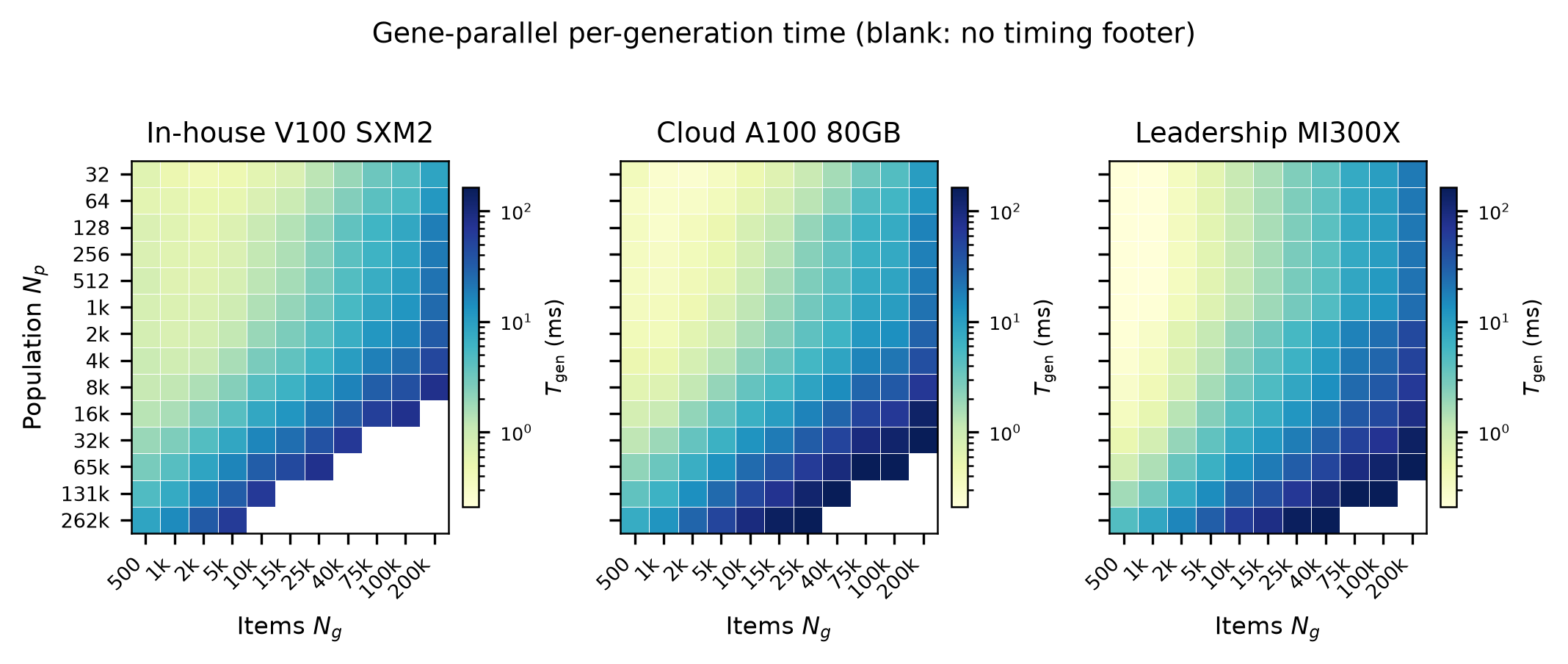}
\caption{Gene-parallel per-generation time on the in-house V100~SXM2, the cloud A100~80GB workstation, and the leadership-class MI300X. Blank cells have no timing footer (out-of-memory or out-of-time) and are not imputed. The colour scale is shared so that occupancy and capacity are visible; each panel is scored only against that home's CPU, not against the other devices.}
\label{fig:tgen_homes}
\end{figure*}

Figure~\ref{fig:tgen_cpu} places that baseline on the $(N_p,N_g)$ plane. One-thread $T_{\mathrm{gen}}$ occupies $10^1$--$10^5$\,ms; the blank corner is the same Large-class capacity/timeout cut used on the devices. Figure~\ref{fig:tgen_homes} is the corresponding gene-parallel field in each deployment. The shared logarithmic colour scale is a \emph{readability} choice so that occupancy (upper-left) and capacity (blank lower-right) can be seen at a glance; it is not a ranking of GPUs. Every populated cell on every device remains in tens of milliseconds until the grid no longer fits.

\subsection{End-to-End Adequacy}
\label{sec:e2e}

A generation is adequate when the device returns a complete footer on a usable inner-loop budget.
Each campaign is scored only against its own CPU (Figs.~\ref{fig:tgen_cpu}--\ref{fig:tgen_homes}; Table~\ref{tab:e2e}).

\paragraph{Tesla V100 SXM2.}
\label{sec:v100}
\begin{itemize}
\item A median gene-parallel generation finishes in $3.84$\,ms against a $575$ ms one-core host step, and at the peak cell $(16384,75000)$ the gap is $58$\,ms versus $37.1$\,s, which is fast enough for interactive generational loops on the V100, without waiting $\approx$0.6 s per step on the host.

\item Geometric-mean $S_{\mathrm{G}}=89.6$ (median $129$) and $S_{72}=12.1$ on the 59 Small pairs show that the win survives even against a fully subscribed socket, so the accelerator is not merely beating an idle host.
\item Chromosome-parallel reaches only $S_{\mathrm{C}}=5.07$ and loses every matched pair (median $\rho=4.81$), which means an outer-loop-only offload buys little more than what $8$--$16$ CPU threads already deliver.
\item At $(32,500)$ the device is slower than one core ($0.49\times$) because four target launches outweigh $0.32$\,ms of useful host work: tiny jobs need a host cutoff, not a faster GPU.
\item The blank lower-right of Fig.~\ref{fig:tgen_homes} (left) is the 16\,GB GDDR5 wall, and the dark cells at $N_p\le 64$ in Fig.~\ref{fig:sp_homes} (left) are occupancy death. Thus, capacity and fill, not algorithm failure, bound this campaign.
\end{itemize}

\paragraph{NVIDIA A100 80GB.}
\label{sec:a100}
\begin{itemize}
\item Median $T_{\mathrm{G}}=3.74$\,ms stays in the same few-millisecond band as V100, but the peak $S_{\mathrm{G}}=817$ at $(65536,100000)$ ($241$\,ms vs $197$\,s) is a Large grid that 16\,GB cannot even store. Ampere's residency allows it to out-shine Volta.
\item $S_{\mathrm{G}}=136$ and $S_{72}=17.2$ confirm a rented workstation is already a production environment against both a weak and a strong host baseline.
\item Gene-parallel still wins every matched pair (median $\rho=5.38$), so the mapping rule travels with the source even when the memory envelope grows.
\item The $(32,500)$ cell remains a loss ($0.81\times$): more SMs and GDDR5 raise the ceiling of what fits, not the floor of what is worth launching.
\item Fig.~\ref{fig:tgen_homes} (center) fills Large cells missing on V100, and Fig.~\ref{fig:sp_homes} (center) lights up for $N_p\ge 8192$. The jump from 16  GB to 80 GB is what turns “out of memory / no timing footer” into “a real timed generation".
\end{itemize}

\paragraph{AMD Instinct MI300X.}
\label{sec:mi300x}
\begin{itemize}
\item Gene-parallel stays in $0.15$--$276$\,ms (median $4.17$\,ms) while chromosome-parallel sits at $17.6$\,ms median, so interactivity holds deep into Large when the gene axis is exposed.
\item $S_{\mathrm{G}}=155$ (peak $1588$ at $(65536,75000)$: $93.1$\,ms vs $148$\,s) and $S_{72}=16.6$ show a leadership node can be filled with the same OpenMP source without a vendor rewrite.
\item The mapping penalty is the widest of the three homes (median $\rho=6.77$) because 304 CUs punish a serial gene walk harder than 80 or 108 SMs---chromosome-parallel never catches up.
\item At $(32,500)$ the device manages only $2.1\times$: the run is launch-bound, so DetermineElite-class kernels and tiny grids remain host work even on CDNA~3.
\item Fig.~\ref{fig:tgen_homes} (right) has the smallest blank corner, and Fig.~\ref{fig:sp_homes} (right) turns yellow-white for $N_p\ge 32768$---capacity is almost gone as a limiter; occupancy is not.
\end{itemize}

\begin{table*}[!t]
\centering
\caption{End-to-end adequacy of OpenMP~5 QIEO. Each column is scored only against that campaign's CPU. No entry is a GPU-to-GPU ratio.
$^{\mathrm{a}}$V100 chromosome-parallel timings exist only on the 59 Small pairs.}
\label{tab:e2e}
\small
\renewcommand{\arraystretch}{1.08}
\begin{tabular}{@{}lccc@{}}
\toprule
 & Tesla V100 SXM2 & NVIDIA A100 80GB & AMD Instinct MI300X \\
\midrule
Median $T_{\mathrm{G}}$ (ms) & $3.84$ & $3.74$ & $4.17$ \\
Median $T_{\mathrm{C}}$ (ms) & $3.38^{\mathrm{a}}$ & $8.72$ & $17.6$ \\
Median $T(1)$ (ms) & $575$ & $575$ & $575$ \\
Gene-parallel $T_{\mathrm{gen}}$ band & $0.43$--$80.4$\,ms & $0.26$--$254$\,ms & $0.15$--$276$\,ms \\
CPU $T_{\mathrm{gen}}$ at Large (ms) & $10^{4}$--$10^{5}$ & $10^{4}$--$10^{5}$ & $10^{4}$--$10^{5}$ \\
\midrule
$S_{\mathrm{G}}$ geometric mean & $89.6$ & $136$ & $155$ \\
$S_{\mathrm{G}}$ median & $129$ & $214$ & $185$ \\
$S_{\mathrm{G}}$ peak & $639$ & $817$ & $1588$ \\
Peak $(N_p,N_g)$ & $(16384,75000)$ & $(65536,100000)$ & $(65536,75000)$ \\
$T_{\mathrm{G}}$ / $T(1)$ at peak & $58$\,ms / $37.1$\,s & $241$\,ms / $197$\,s & $93.1$\,ms / $148$\,s \\
\midrule
$S_{72}$ geometric mean & $12.1$ & $17.2$ & $16.6$ \\
$S_{72}$ median & $14.5$ & $24.3$ & $20.5$ \\
$S_{72}$ maximum & $22.3$ & $36.3$ & $62.6$ \\
$S_{72}$ pair set & 59 Small & 59 Small & 59 Small \\
\midrule
$S_{\mathrm{C}}$ geometric mean & $5.07$ & $15.4$ & $21.9$ \\
Median $\rho=T_{\mathrm{C}}/T_{\mathrm{G}}$ & $4.81$ & $5.38$ & $6.77$ \\
Geometric-mean $\rho$ & $4.16$ & $4.89$ & $5.73$ \\
Gene faster than chromosome & $100\%$ & $100\%$ & $100\%$ \\
\midrule
$S_{\mathrm{G}}(32,500)$ & $0.49\times$ & $0.81\times$ & $2.1\times$ \\
Host work at $(32,500)$ & $0.32$\,ms & $0.32$\,ms & $0.32$\,ms \\
Occupancy-dark $N_p$ & $\le 64$ & $\le 64$ & $\le 64$ \\
Large-speedup region & $N_p\ge 4096$, $N_g\ge 15000$ & $N_p\ge 8192$ & $N_p\ge 32768$, $N_g\ge 5000$ \\
\bottomrule
\end{tabular}
\end{table*}

Gene-parallel offload moves a $\sim$0.6\,s host step into a few milliseconds on every GPU and still beats 72 host threads by $12$--$17\times$ (Table~\ref{tab:e2e}).
Chromosome-parallel never wins a matched pair ($\rho>1$ everywhere; Table~\ref{tab:rho}).
Tiny grids belong on the host; capacity, not the algorithm, is what 16\,GB truncates.

\subsection{Occupancy Challenges}
\label{sec:occupancy}

Occupancy is whether the launch feeds the mesh.
Chromosome-only mapping starves every GPU; the fill constant simply tracks SM or CU count.

\begin{figure*}[!t]
\centering
\includegraphics[width=0.98\textwidth]{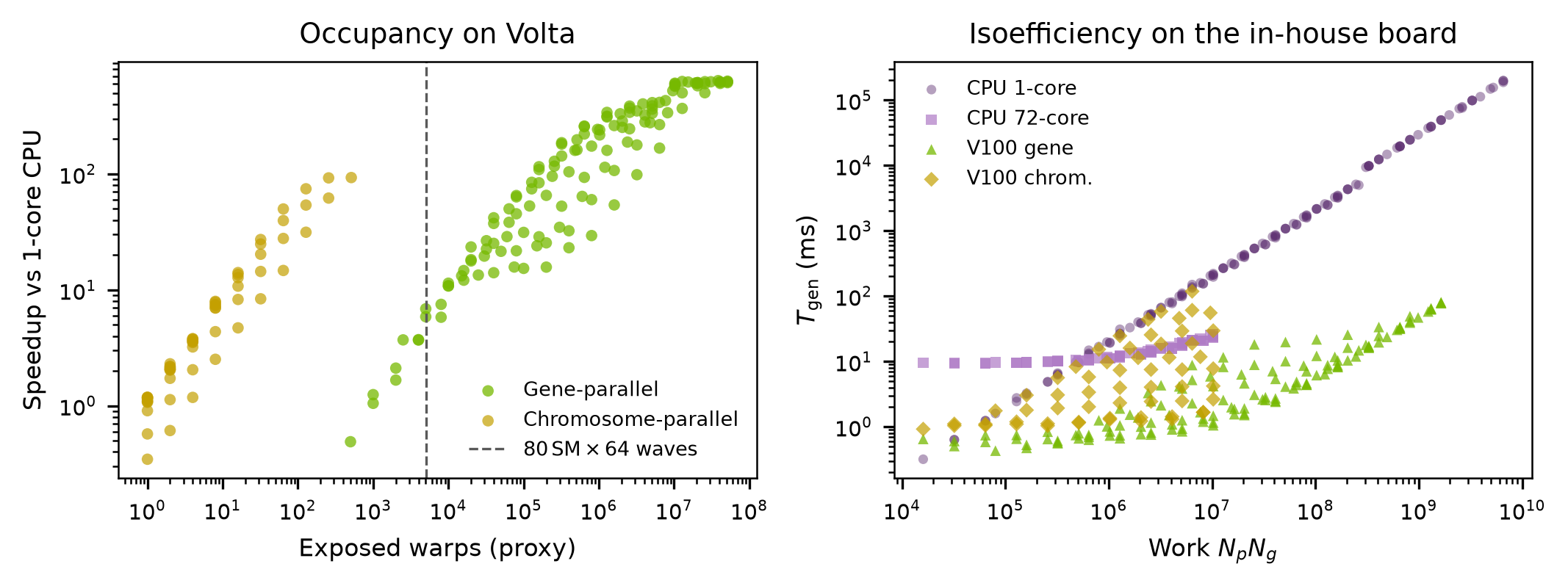}
\caption{Left: V100 speedup versus an exposed-warp proxy. The dashed line is $80~\mathrm{SM}\times 64$ waves. Right: isoefficiency $T_{\mathrm{gen}}$ versus $W=N_p N_g$ on the in-house board.}
\label{fig:v100_occ}
\end{figure*}

\paragraph{Tesla V100 SXM2.}
\begin{itemize}
\item A chromosome-parallel kernel with $N_p=32$ feeds a single warp to 80 SMs, while the same population at $N_g=500$ under gene-parallel still supplies only 500 warps against a generous $\sim 5120$-warp ceiling---Volta is starved long before it is compute-bound.
\item Fig.~\ref{fig:v100_occ} (left) makes the mapping choice visible: chromosome points live left of the warp-proxy line and rarely exceed $10\times$, whereas gene-parallel climbs toward $600\times$ once the mesh is fed.
\item The practical rule is therefore a host cutoff $N_p N_g < 8\cdot 80\cdot 256$ and, on device, \texttt{num\_teams(80)} with \texttt{thread\_limit(128)} or \texttt{(256)} so the launch width matches the SM count rather than the outer loop.
\item Isoefficiency against $W=N_p N_g$ (Fig.~\ref{fig:v100_occ}, right) sits one to two decades below the serial CPU, with branches at constant $W$: skinny populations of long chromosomes are not interchangeable with fat populations of short ones.
\end{itemize}

\paragraph{NVIDIA A100 80GB.}
\begin{itemize}
\item The same occupancy arithmetic applies on a wider mesh: $N_p=32$ is still one warp for 108 SMs, and $N_g=500$ gene-parallel remains below $108\times 64=6912$ concurrent warps, so Ampere raises the fill requirement rather than removing it.
\item Moving the host cutoff to $N_p N_g < 8\cdot 108\cdot 256$ and launching with \texttt{num\_teams(108)} (and \texttt{thread\_limit(256)} or $512$ once 164\,KiB tiles are live) is how the extra SMs become useful instead of idle.
\item MIG slicing reduces SM count and L2 and is strictly harmful for a single dense generation: partitioning the board undoes the occupancy the mapping just earned.
\item Tensor Cores do not accelerate double-precision qubit rotations or integer knapsack products; forcing TF32 would change the search, not repair under-fill.
\end{itemize}

\begin{figure}[!t]
\centering
\includegraphics[width=\columnwidth]{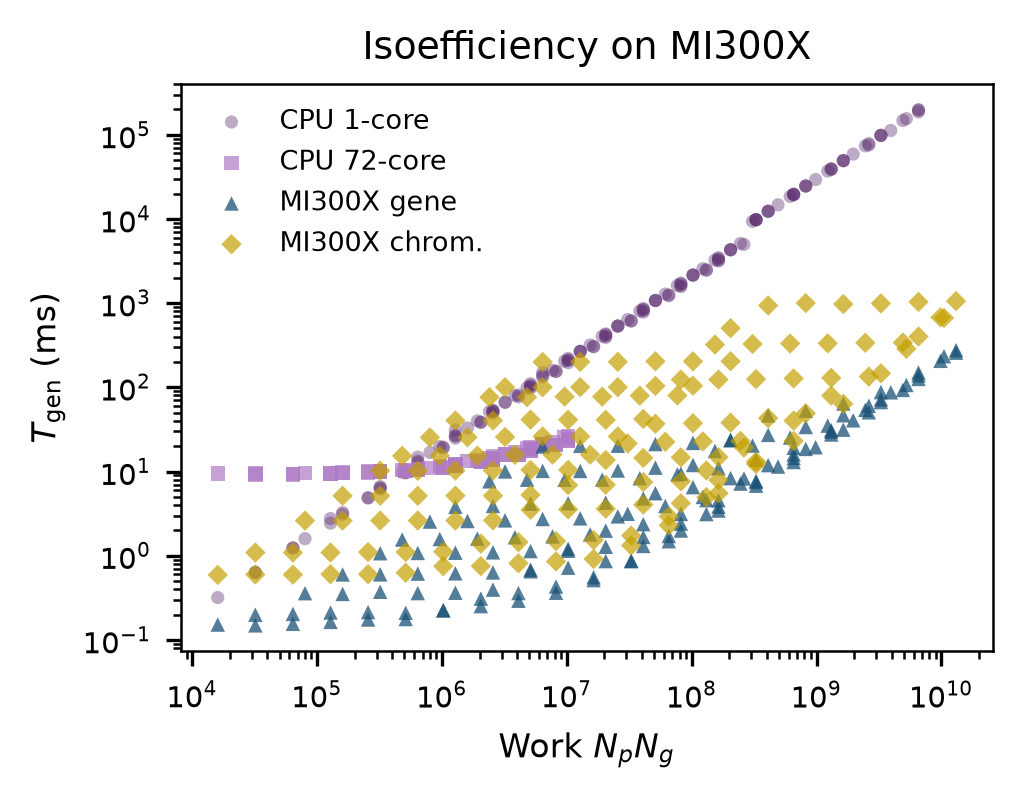}
\caption{Isoefficiency on the leadership node. Branches at constant $W=N_p N_g$ show that the aspect ratio of the qubit grid, not merely its area, controls device time.}
\label{fig:mi_iso}
\end{figure}

\paragraph{AMD Instinct MI300X.}
\begin{itemize}
\item With 64-lane wavefronts across eight XCDs, $N_p=32$ chromosome-parallel supplies half a wavefront and leaves seven dies idle---the mesh-shaped cost of reading the flowchart literally~\cite{amdmi300x}.
\item Gene-parallel at $(32,500)$ still yields only $\sim 250$ wavefronts against an order of $10^5$ threads needed to feed 304 CUs; QIP crosses that line once $N_p N_g\gtrsim 10^5$, which is why Small grids remain launch-bound here as elsewhere.
\item The darkest $\rho$ cells sit at mid $N_p$ and large $N_g$, where each team still walks a long serial gene loop; as $N_p$ grows, $\rho$ falls toward $2$--$3$ but never crosses (Fig.~\ref{fig:sp_homes}, right)---there is no late crossover that would justify chromosome-parallel on CDNA~3.
\item The host cutoff becomes $N_p N_g < 8\cdot 304\cdot 256$, and device launches need a team count on the order of 304 CUs (or 38 per XCD); a single-digit team count wastes the MCM.
\item Isoefficiency (Fig.~\ref{fig:mi_iso}) again branches at constant $W$: Evaluation reductions and wavefront occupancy depend on the aspect ratio $N_p:N_g$, not merely on work volume.
\end{itemize}

Eighty SMs, 108 SMs, and 304 CUs are three constants in the same fill rule: expose the gene axis, or leave the mesh idle.
If $N_p N_g$ is below about one 256-thread block per SM/CU, keep the generation on the host (Section~\ref{sec:play}).
Do not MIG-slice a single dense job.

\subsection{Memory Management}
\label{sec:memory}

$Q$ is the large resident object, while the 0/1 knapsack evaluator-specific values adn weights, $(v,w)$, are small and should be stored in the fastest accessible memory resource. As the problem size increases, the cut-off change with the memory capacity available, but the aforementioned placements priority remains unchanged.

\begin{figure}[!t]
\centering
\begin{tikzpicture}[
  font=\scriptsize\sffamily,
  box/.style={draw, rounded corners=1.2pt, align=center, inner sep=3pt, minimum width=2.35cm},
  arr/.style={-{Latex}, thick}
]
\node[box, fill=black!12] (reg) {Registers / warp\\[-1pt]{\tiny accumulators, shuffles}};
\node[box, fill=black!8, below=3.2mm of reg] (sh) {Shared $\approx 48$\,KiB\\[-1pt]{\tiny $(v,w)$ if $N_g\le 4915$}\\[-1pt]{\tiny else tiles of 128}};
\node[box, fill=black!4, below=3.2mm of sh] (cst) {Constant 64\,KiB\\[-1pt]{\tiny $(v,w)$ if $N_g\le 6553$}};
\node[box, fill=white, below=3.2mm of cst] (hbm) {GDDR5 16\,GB\\[-1pt]{\tiny $Q$, $x$ always}\\[-1pt]{\tiny $(v,w)$ if $N_g>6553$}};
\draw[arr] (reg) -- (sh);
\draw[arr] (sh) -- (cst);
\draw[arr] (cst) -- (hbm);
\end{tikzpicture}
\caption{V100 memory layout for one QIEO generation. Place $(v,w)$ as high as they fit; keep $Q$ resident in GDDR5 across generations.}
\label{fig:v100_layout}
\end{figure}
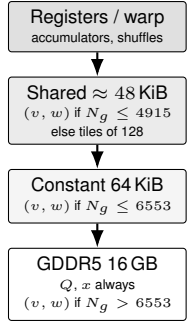

\begin{figure*}[!t]
\centering
\includegraphics[width=0.98\textwidth]{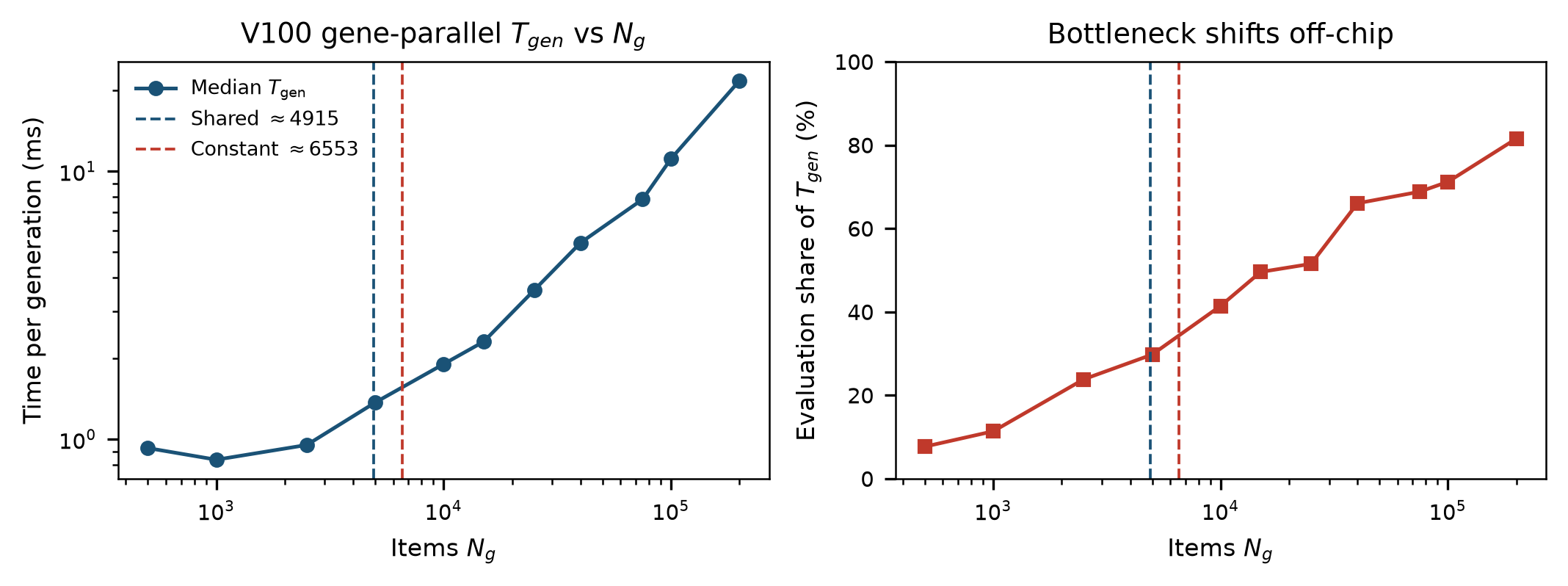}
\caption{V100 gene-parallel memory cliffs. Left: median $T_{\mathrm{gen}}$ versus $N_g$, with shared- and constant-memory capacities. Right: Evaluation's share of the generation; the bottleneck leaves on-chip memory with $N_g$.}
\label{fig:v100_mem}
\end{figure*}

\begin{figure}[!t]
\centering
\includegraphics[width=\columnwidth]{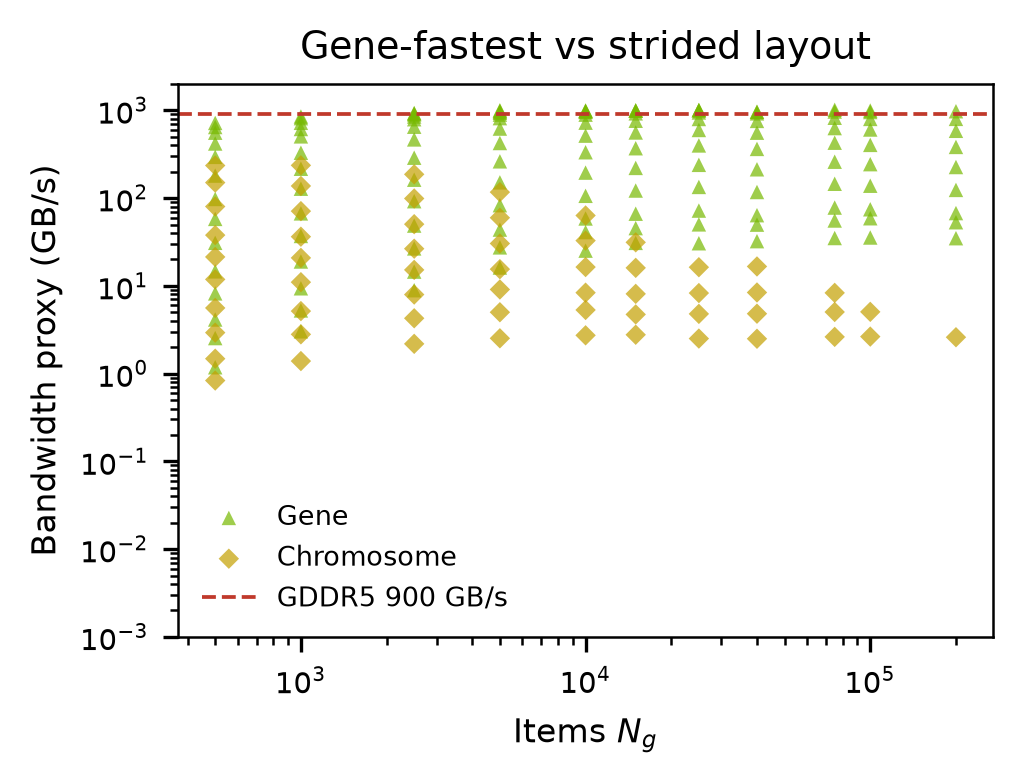}
\caption{Streaming-bandwidth proxy versus $N_g$ on V100. Gene-parallel approaches the 900\,GB/s GDDR5 roof; chromosome-parallel remains one to two decades below because the inner loop is uncoalesced.}
\label{fig:v100_bw}
\end{figure}

\paragraph{Tesla V100 SXM2.}
\begin{itemize}
\item Fig.~\ref{fig:v100_layout} and Eqs.~\eqref{eq:nshv}--\eqref{eq:nconst} fix the Volta placement. They suggest keeping $Q$ in GDDR5, and routing $(v,w)$ through shared memory, then constant memory, then 128-item tiles once both on-chip homes are exceeded. Past those capacity cutoffs, Evaluation must stream $(v,w)$ from GDDR5 instead of reading them from on-chip memory, so its cost and share of $T_{gen}$ rise sharply.

\item Gene-fastest layout of $Q$ lets a 32-thread warp issue a coalesced 128-byte transaction, whereas a chromosome-major inner loop strides by $N_g$ and misses that path, which is why chromosome-parallel never approaches the memory roof.

\item Below $\sim 2$\,k items, $T_{\mathrm{gen}}$ is almost flat indicating launch domination. Past the shared/constant capacity marks, Evaluation streams from GDDR5 and its share of the generation climbs from $0.077$ to $0.816$ (Fig.~\ref{fig:v100_mem}. As Table~\ref{tab:evaln})indicates, its memory hierarchy and not arithmetic intensity that sets the Large-class bottleneck.

\item Fig.~\ref{fig:v100_bw} confirms the layout claim: gene-parallel approaches the $900$\,GB/s GDDR5 roof while chromosome-parallel stays one to two decades below.

\item Sixteen gigabytes is a hard residency wall for double-precision grids, so the first HtoD of $Q$ should use pinned staging and every later generation should reuse the device copy.
\end{itemize}

\begin{figure*}[!t]
\centering
\includegraphics[width=0.98\textwidth]{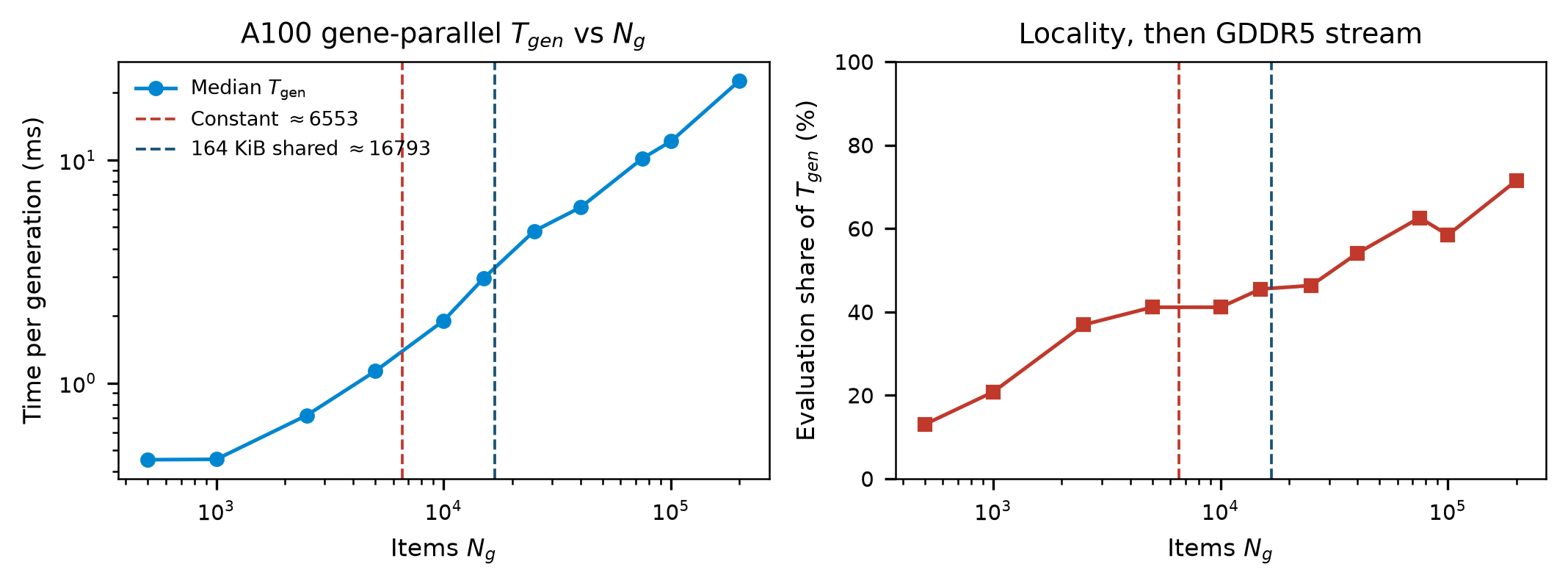}
\caption{A100 gene-parallel memory cliffs. Left: median $T_{\mathrm{gen}}$ versus $N_g$, with constant-memory and 164\,KiB shared-memory capacities. Right: Evaluation's share of the generation. Tiles of 256--512 items are legal past the Volta shared-memory cliff.}
\label{fig:a100_mem}
\end{figure*}

\paragraph{NVIDIA A100 80GB.}
\begin{itemize}
\item Constant memory is still 64\,KiB~\eqref{eq:nconst}, but shared memory is $3.4\times$ the Volta default~\eqref{eq:nsha}, so 256--512-item tiles that would evict on V100 are legal here.
\item Item tables at $N_g=200000$ (1.6\,MB) fit comfortably in the 40\,MB L2, while the 80\,GB GDDR5 exists to hold is the qubit grid itself, which transforms Large cases from a capacity failure into a locality problem.

\item Once Evaluation must stream $(v,w)$ from GDDR5, median $T_{\mathrm{gen}}$ rises and Evaluation's share grows from $0.130$ to $0.715$ (Fig.~\ref{fig:a100_mem}; Table~\ref{tab:evaln}), even while the tables still fit in L2, because the \emph{grid} does not.
\item CIP's share grows when $Q$ leaves the L2 window: elite broadcast becomes a GDDR5 walk, so pinning $(v,w)$ and the elite with an access-policy window protects the only reuse in the generation.
\item \texttt{cp.async} double-buffered tiles, that L2 window, and a single \texttt{target data} region around the generational loop are how Ampere extracts the last factor; they are not prerequisites for the code to compile or to beat the host.
\end{itemize}

\paragraph{AMD Instinct MI300X.}
\begin{itemize}
\item Weights and values of $8N_g$ bytes (1.6\,MB at $N_g=200000$) fit in Infinity Cache and should be \texttt{map(to:)} once, so the leadership memory hierarchy's useful work is keeping $(v,w)$ hot rather than chasing HBM capacity.
\item At $(262144,200000)$ the qubit grid is tens of gigabytes and is why a few Large cells still lack footers. Even 192\,GB HBM3 proves to be insufficient when both axes grow together.

\item Gene-fastest layout lets 64-lane wavefronts coalesce on HBM3, and team-private elite caches cut CIP's broadcast traffic. Layout mistakes here waste bandwidth the same way they do on Volta and Ampere.

\item Fusing QIP+observe and Evaluation, hosting DetermineElite, and keeping CIP as a separate region cuts four implicit device barriers to two. This is a critical improvement since launch tax matters on a multi-die GPU even when capacity does not.
\end{itemize}

The key learnings are to keep $Q$ resident, store $(v,w)$ as high as they fit (shared $\to$ constant $\to$ tiled, or Infinity Cache on MI300X). Past those cliffs Evaluation streams from device memory and owns the generation time (Table~\ref{tab:evaln}). Also, keep $Q$ in gene-fastest (coalesced) layout on the device, and transfer it from the host only once for the whole run.

\subsection{Kernel-Resolved Picture}
\label{sec:kernels}

The four operations of QIEO can be categorized as, QIP and CIP are maps; Evaluation is a segmented reduction; DetermineElite is a scalar minimum.
The question is which stage owns the millisecond budget (Tables~\ref{tab:ksp}--\ref{tab:evaln}).

\paragraph{Tesla V100 SXM2.}
\begin{itemize}
\item Gene-parallel QIP ($537\times$) and CIP ($165\times$) are the easy wins, while Evaluation stalls near $12\times$ because it is a segmented reduction that cannot use the same element-wise map.

\item DetermineElite at $0.060\times$ is a $17\times$ slowdown. Microseconds of host arithmetic become a device launch plus a DtoH of one scalar, so selection does not belong in the offload pipeline.

\item On the CPU, QIP is 77\% of $T_{\mathrm{gen}}$; gene-parallel flips the profile to Evaluation 29\% overall and 82\% at $N_g=200000$, whereas chromosome-parallel retains a CPU-like 57\% QIP (Table~\ref{tab:fracall}),so which kernel dominates $T_{\mathrm{gen}}$ is set by gene- vs chromosome-parallel offload, not by which GPU runs it.
\end{itemize}

\paragraph{NVIDIA A100 80GB.}
\begin{itemize}
\item The same ranking holds with larger absolute factors: QIP $678\times$ and CIP $219\times$ carry the generation, Evaluation caps near $16\times$, and DetermineElite remains a $0.063\times$ tax (Table~\ref{tab:ksp}).

\item Evaluation's share is 29\% overall and 71\% at $N_g=200000$, while chromosome-parallel still looks like a CPU run (58\% QIP), so Ampere extras that only polish Evaluation cannot rescue a chromosome-only mapping.

\item Larger tiles, an L2 window, and \texttt{cp.async} squeeze the reduction that owns Large-class time; they do not replace \texttt{collapse(2)} on the gene axis.
\end{itemize}

\paragraph{AMD Instinct MI300X.}
\begin{itemize}
\item Gene-parallel QIP reaches $1010\times$ and CIP $377\times$, but Evaluation only $10\times$ and DetermineElite $0.105\times$ (Table~\ref{tab:ksp}): CDNA~3 has already made the maps cheap, so further device work belongs in the reduction.
\item Evaluation owns 64\% of gene-parallel time and rises from 49\% to 87\% with $N_g$, while QIP falls to 15\%---speeding PopulationReset after that point has diminishing returns.
\item Chromosome-parallel keeps a CPU-shaped profile on HBM3 (QIP 64\%, CIP 26\%), proving again that the mapping,not the Mulit-Chip Module(MCM), decides which kernel is critical (Table~\ref{tab:fracall}).

\item The profitable moves are a better segmented reduction for~\eqref{eq:fitness}, a host DetermineElite after Evaluation, and a register-cached elite row to close CIP's remaining gap to QIP.
\end{itemize}

The key learnings are, QIP and CIP are solved on every GPU; Evaluation is the portable ceiling ($10$--$16\times$) and becomes the majority of gene-parallel time at large $N_g$ (Table~\ref{tab:evaln}).
DetermineElite is a launch-tax liability everywhere and is best to leave it on the host. Spend the tuning budget on the segmented reduction, not on another pragma for QIP.

\section{Discussions}
\label{sec:three}

The three accelerators are not interchangeable silicon with different clock rates. They are three answers to three operational questions: can the laboratory already run QIEO; can a rented workstation hold the grids the laboratory cannot; and can a leadership node fill an allocation without a vendor rewrite. Architecture sets the constraint, performance is judged against the host in that chassis, and the role follows from both (Table~\ref{tab:hw}, Figs.~\ref{fig:tgen_homes}--\ref{fig:sp_homes}).

\subsection{Architecture}

Volta, Ampere, and CDNA~3 expose the same $N_p\times N_g$ qubit grid to very different schedulers and memory systems.

Tesla V100~SXM2 is a single-die Volta GPU: 80 SMs, 32-thread warps, 16\,GB of GDDR5 at $\approx 900$\,GB/s, and a 64\,KiB constant cache that still matters~\cite{volta2018,v100ds}. For QIEO the architectural fact is scarcity. Eighty SMs starve unless the gene axis is exposed; 16\,GB is a hard residency ceiling; constant and $\approx 48$\,KiB shared memory are the only on-chip homes for $(v,w)$ before Evaluation becomes a GDDR5 stream. There is no large L2 window and no asynchronous copy pipeline. What Volta does offer is a mature OpenMP offload path and a board that a laboratory typically already owns.

A100~80GB is Ampere with the same warp size but a different memory thesis~\cite{a100wp,a100ds}. One hundred eight SMs raise the occupancy floor; 80\,GB of GDDR5 at $\approx 2.0$\,TB/s changes Large from ``does not fit'' to ``must be placed''; 164\,KiB of shared memory legalizes tiles that evict on Volta; 40\,MB of L2 with an access-policy window is enough to pin the item tables and the elite chromosome while the grid streams. \texttt{cp.async} is the Ampere-specific way to hide that stream. Tensor Cores and MIG are architectural features that this workload should ignore, since rotations and integer knapsack products are not tensor arithmetic, and slicing the SM/L2 budget is strictly harmful for one dense generation.

MI300X is a different machine class~\cite{amdmi300x}. Eight XCDs, 304 compute units, 64-lane wavefronts, 192\,GB of HBM3 at 5.3\,TB/s, and a 256\,MB Infinity Cache shared across the MCM. The occupancy problem is now mesh-shaped: a chromosome-parallel kernel that under-fills one Volta SM under-fills an entire XCD fabric. The capacity problem is almost gone for this archive, since item tables of $1.6$\,MB sit inside Infinity Cache, and only the most extreme qubit grids exhaust 192\,GB. The compiler is ROCm OpenMP, not nvc++, so portability is a language contract rather than a vendor lock. Cross-XCD reductions, however, make tiny kernels such as DetermineElite even less defensible than on a single NVIDIA die.

The architectural invariant is coarser than any of those constants. QIEO is a streaming occupancy problem on every card. Warps or wavefronts must be fed from the gene axis; $(v,w)$ must live in whichever on-chip structure the GPU provides; $Q$ must remain resident. What changes is which structure, and how soon the grid no longer fits.

\subsection{Performance}

``Performing well'' is not a single ranking. It is adequacy against the CPU that came with the board, plus the ability to remain interactive as $(N_p,N_g)$ grows (Table~\ref{tab:e2e}).

On the in-house V100, gene-parallel OpenMP already moves the generation off the human-timescale host: median $3.84$\,ms versus $575$\,ms on one core, geometric-mean $90\times$ versus that core and $12\times$ versus 72 host threads. The peak $639\times$ at $(16384,75000)$ is occupancy earned, not clock rate. The same board, mapped over chromosomes, returns only $5\times$, inside the host's own 8--16-thread band, so a poorly oriented offload spends Volta to buy what DRAM-bound OpenMP already delivered. The blank lower-right of Fig.~\ref{fig:tgen_homes} (left) is the performance statement the architecture predicted: 16\,GB is the campaign limit, not the algorithm.

On the cloud A100 the median generation is still a few milliseconds ($3.74$\,ms), but the \emph{populated} grid is wider. Geometric-mean $136\times$ versus one core and $17\times$ versus 72 threads, with a peak of $817\times$ at $(65536,100000)$, mean that a rented workstation is already a production environment. Ampere does not repeal launch tax, i.e., $(32,500)$ is still $0.81\times$ versus one core. Instead it raises the ceiling. Cells that have no footer on V100 produce timings here because 80\,GB holds the grid. Kernel-resolved, QIP and CIP are solved ($678\times$, $219\times$); Evaluation at $16\times$ is the new critical path, which is why \texttt{cp.async} tiles and an L2 window are how this GPU extracts the last factor rather than how it becomes viable.

On MI300X the generation remains in $0.15$--$126$\,ms (median $4.17$\,ms) deep into Large. Geometric-mean $155\times$ versus one core and $16.6\times$ versus 72 threads, with a peak of $1588\times$, show that a leadership node can be filled with the same source. The mapping penalty is the largest of the three homes (median $\rho=6.77$) because 304 CUs punish a serial gene walk more severely than 80 SMs do; there is still no crossover. Evaluation owns 64--87\% of gene-parallel time: CDNA~3 has already made QIP cheap, so further device work belongs in the segmented reduction, not in another pragma on PopulationReset. The smallest blank corner in Fig.~\ref{fig:tgen_homes} is the leadership-class role in one picture.

Two performance facts travel with the source rather than with the card. First, gene-parallel wins every matched pair on every device; chromosome-parallel never inverts the ranking. Second, DetermineElite is a slowdown everywhere ($0.06\times$--$0.11\times$). Those are not GPU results. They are the portable contract succeeding or failing independently of Volta, Ampere, or CDNA~3.

\subsection{Roles}

The right GPU ecosystem is the one that matches the job, not the one with the largest $S_{\mathrm{G}}$.

The V100 is the development and regression home. It sits in the rack, compiles with a toolchain the laboratory owns, and is fast enough for Small and Medium campaigns once \texttt{collapse(2)} is on. Its job is to make QIEO a daily tool rather than a reservation. When a Large grid refuses to produce a footer, the correct response is to move the run, not to rewrite the algorithm.

The A100~80GB is the campaign home. It is what a cloud catalog offers when the laboratory has outgrown 16\,GB but does not yet need a center allocation. Its job is residency and locality: keep $Q$ on the device, place $(v,w)$ in constant memory, a 164\,KiB tile, or an L2 window, and do not MIG-slice a single generation. A workstation that holds $(65536,100000)$ in $241$\,ms/gen is doing the work the V100 physically cannot.

The MI300X is the allocation home. Its job is to absorb the grids and the CU count that exist so that a leadership node is not idle. Infinity Cache swallows the item tables; 192\,GB swallows almost every qubit grid in this archive; 304 CUs punish any mapping that is not gene-parallel. Compiling with ROCm rather than nvc++ is the point of the portable source.

A production workflow therefore moves the \emph{same binary contract} along Fig.~\ref{fig:ecosystem}: prototype and nightly tests on V100, Large interactive campaigns on A100, and the extreme $(N_p,N_g)$ corner on MI300X. Architecture explains why each move is necessary; the speedups against each home's CPU explain why none of the three is a correctness vehicle.

\subsection{Cross-Cutting Evidence}

\begin{figure*}[!t]
\centering
\includegraphics[width=0.98\textwidth]{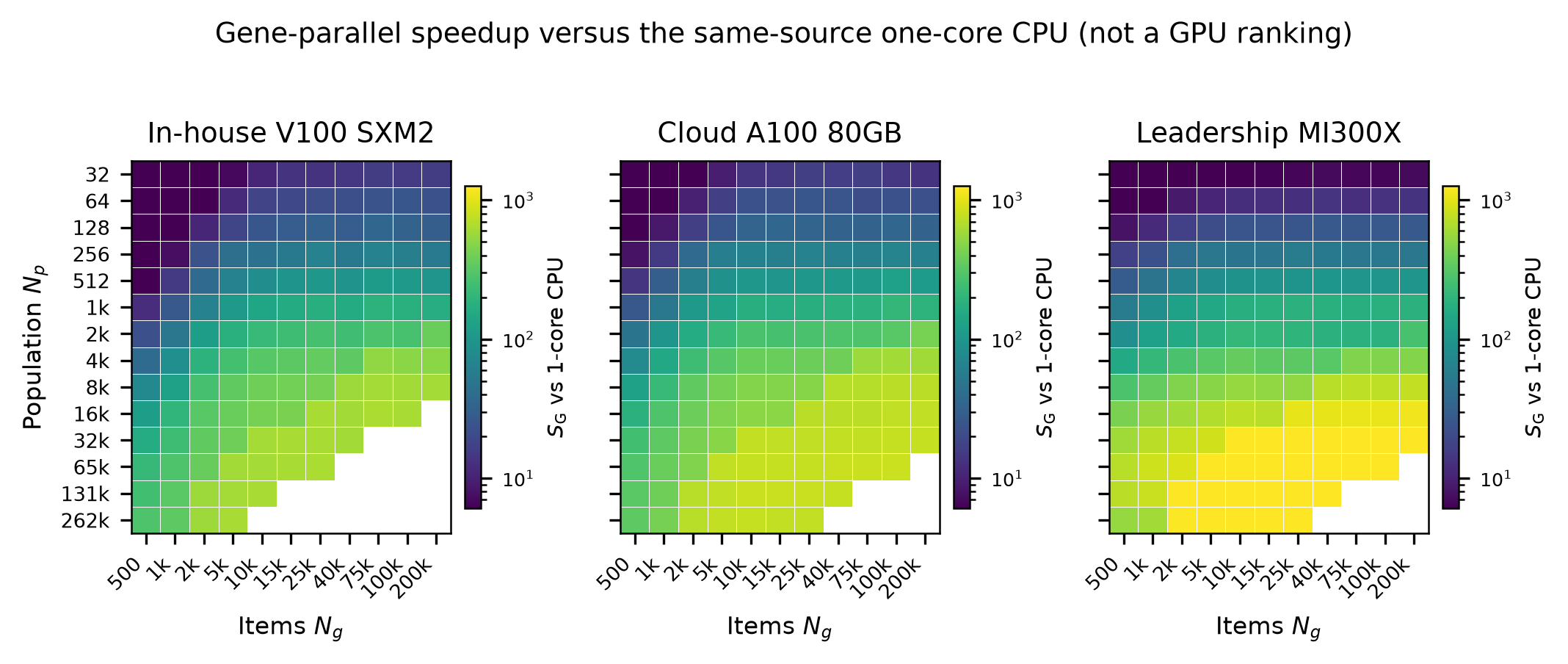}
\caption{Gene-parallel speedup $S_{\mathrm{G}}=T(1)/T_{\mathrm{G}}$ versus the same-source one-core CPU. Dark cells at small $N_p$ are occupancy-limited; yellow cells are production-scale grids. Blank cells have no matched device footer. No panel is a ratio of one GPU to another.}
\label{fig:sp_homes}
\end{figure*}

The three campaigns were scored independently. They nevertheless agree on a small set of facts that do not require one GPU to be compared with another. Figure~\ref{fig:sp_homes} restates adequacy as a speedup field. Each panel is $T(1)/T_{\mathrm{G}}$ against that campaign's CPU. Tables~\ref{tab:e2e}--\ref{tab:evaln} collect the same facts numerically so that a practitioner moving from the laboratory to the cloud to a leadership node can see what is portable and what must be retuned.

\begin{table}[!t]
\centering
\caption{Mapping penalty $\rho=T_{\mathrm{C}}/T_{\mathrm{G}}$ and the $(32,500)$ launch-tax case. Gene-parallel wins every matched pair in every environment.}
\label{tab:rho}
\begin{tabular}{lccc}
\toprule
Deployment & Median $\rho$ & Pairs with $\rho>1$ & $S_{\mathrm{G}}(32,500)$ \\
\midrule
In-house V100 & 4.81 & 100\% & $0.49\times$ \\
Cloud A100 & 5.38 & 100\% & $0.81\times$ \\
Leadership MI300X & 6.77 & 100\% & $2.1\times$ \\
\bottomrule
\end{tabular}
\end{table}

\begin{table*}[!t]
\centering
\caption{Geometric-mean kernel speedup versus one CPU core, matched on $(N_p,N_g)$.}
\label{tab:ksp}
\begin{tabular}{lcccccc}
\toprule
 & \multicolumn{2}{c}{In-house V100} & \multicolumn{2}{c}{Cloud A100} & \multicolumn{2}{c}{Leadership MI300X} \\
\cmidrule(lr){2-3}\cmidrule(lr){4-5}\cmidrule(lr){6-7}
Kernel & Gene & Chrom. & Gene & Chrom. & Gene & Chrom. \\
\midrule
Per-generation total & 89.6 & 5.07 & 136.0 & 15.4 & 155.0 & 21.9 \\
QIP      & 536.7 & 8.94 & 677.6 & 25.5 & 1010.3 & 28.0 \\
Evaluation           & 11.9  & 2.17 & 15.6  & 7.5  & 10.2  & 7.8 \\
DetermineElite       & 0.060 & 0.023 & 0.063 & 0.041 & 0.105 & 0.093 \\
CIP                  & 165.3 & 3.72 & 218.6 & 10.8 & 376.6 & 14.8 \\
\bottomrule
\end{tabular}
\end{table*}

\begin{table*}[!t]
\centering
\caption{Median kernel fractions of $T_{\mathrm{gen}}$. CPU numbers are one-thread baselines shared by the archives.}
\label{tab:fracall}
\small
\begin{tabular}{lccccccc}
\toprule
Kernel & CPU $p{=}1$ & V100 gene & V100 chrom. & A100 gene & A100 chrom. & MI300X gene & MI300X chrom. \\
\midrule
QIP & 0.772 & 0.171 & 0.569 & 0.208 & 0.578 & 0.154 & 0.635 \\
Evaluation      & 0.037 & 0.291 & 0.101 & 0.286 & 0.090 & 0.644 & 0.105 \\
DetermineElite  & $<10^{-5}$ & 0.012 & 0.004 & 0.012 & 0.004 & 0.007 & 0.001 \\
CIP             & 0.190 & 0.116 & 0.235 & 0.143 & 0.250 & 0.105 & 0.256 \\
\bottomrule
\end{tabular}
\end{table*}

\begin{table}[!t]
\centering
\caption{Median Evaluation fraction of gene-parallel $T_{\mathrm{gen}}$ versus item count. Intermediate MI300X bins were reported as a monotone rise from 0.49 to 0.87.}
\label{tab:evaln}
\begin{tabular}{cccc}
\toprule
$N_g$ & V100 & A100 & MI300X \\
\midrule
500 & 0.077 & 0.130 & 0.49 \\
2500 & 0.238 & 0.369 & --- \\
5000 & 0.298 & 0.411 & --- \\
10000 & 0.414 & 0.411 & --- \\
40000 & 0.660 & 0.541 & --- \\
200000 & 0.816 & 0.715 & 0.87 \\
\bottomrule
\end{tabular}
\end{table}

Four observations are invariant.

\paragraph{Gene-parallel mapping is not optional.}
In every environment, $\rho>1$ on 100\% of matched pairs (Table~\ref{tab:rho}). The median penalty for reading the flowchart literally, i.e. parallelize chromosomes, is $4.8\times$ on the in-house board, $5.4\times$ on the cloud workstation, and $6.8\times$ on the leadership node. That penalty is larger, in every campaign, than the entire 72-thread CPU gain. Chromosome-parallel offload retains a CPU-shaped QIP dominance (57--64\% of $T_{\mathrm{gen}}$) on silicon that is designed to stream a grid.

\paragraph{The dense kernels are solved; Evaluation is not.}
QIP and CIP accelerate by $10^2$--$10^3\times$ under gene-parallel mapping in every environment (Table~\ref{tab:ksp}). Evaluation saturates near $10$--$16\times$ because it is a segmented reduction of length $N_g$ per chromosome. On gene-parallel devices Evaluation's share of $T_{\mathrm{gen}}$ grows with $N_g$ and becomes the majority at Large (Table~\ref{tab:evaln}). Further work on QIP has diminishing returns; the profitable device kernel is a warp- or wavefront-level reduction for~\eqref{eq:fitness}.

\paragraph{DetermineElite is a launch-tax liability.}
Geometric-mean DetermineElite ``speedups'' of $0.06\times$ (V100, A100) and $0.11\times$ (MI300X) are slowdowns of one order of magnitude. An $O(N_p)$ minimum over doubles does not amortize a target region. Host-side DetermineElite after Evaluation supplies CIP with a known elite. 

\paragraph{Tiny jobs belong on the host.}
The $(32,500)$ cell is slower than one CPU core on V100 and A100, and only $2.1\times$ on MI300X. Four target regions times launch latency exceeds a few hundred microseconds of useful CPU work. A production runtime should skip offload when $N_p N_g$ is below a few blocks per SM or CU (Section~\ref{sec:play}).

Convergence generations depend on $N_g$, not on the device or the mapping. Mapping changes wall time per step, not the search. Solution quality remains a property of QIEO, not of OpenMP.

\section{Recommendations}
\label{sec:play}

The measurements support a two-layer recipe: portable decisions that travel with the source, and environment-specific constants that travel with the machine.

\subsection{Portable layer (every environment)}

\paragraph{P0 --- Host versus device.}
If $N_p N_g$ is below one well-filled wave per compute unit (approximately $10^5$ on V100, $2\cdot 10^5$ on A100, $10^5$ on MI300X), stay on the CPU. The $(32,500)$ cell is the occupancy math, not an outlier.

\paragraph{P1 --- Collapse the qubit grid.}
\texttt{collapse(2)} on QIP and CIP. This single pragma is the difference between a $5$--$22\times$ chromosome-parallel curiosity and a $90$--$155\times$ production mapping versus one core.

\paragraph{P2 --- Persist the working set.}
One outer \texttt{target data} region across generations. \texttt{map(to:)} the read-only $(v,w)$ once. Pinned host staging for the initial copy of $Q$. Gene-fastest layout so that warps or wavefronts coalesce.

\paragraph{P3 --- Reduce Evaluation in the warp or wavefront.}
One accumulator per thread over a strided chunk of items, shuffle-reduce, one store per chromosome. OpenMP~5 \texttt{reduction} on an array of length $N_p$ can lower to this if the compiler sees the inner loop; if it does not, the region should be an explicit local accumulator or a small vendor kernel. This is the kernel that owns most of Large-class $T_{\mathrm{gen}}$.

\paragraph{P4 --- Remove DetermineElite from the device pipeline.}
Host reduction after Evaluation and before CIP, or fuse a device-wide reduce into Evaluation's last write so the OpenMP runtime is not entered between them. Four target regions per generation is four times the launch tax that Small cases cannot afford and that Large cases still pay at the percent level.

\paragraph{P5 --- Do not chase 72 host threads.}
Eight to sixteen CPU cores are the honest baseline. Seventy-two threads buy a median $2.62\times$ and a confused speedup narrative.

\paragraph{P6 --- Fuse barriers when the dependence allows.}
QIP and observation write $Q$ and $x$ at identical $(i,j)$ indices. Cache the elite chromosome in shared memory (or LDS) at the start of CIP. A fused QIP+observe kernel, a fused Evaluation, a host DetermineElite, and CIP reduce the pipeline from four device launches to two.

\subsection{In-house layer (V100 SXM2)}

Place $(v,w)$ in constant memory for $N_g\le 6553$, in shared memory for $N_g\le 4915$, and in shared tiles of 128 items thereafter. Launch with \texttt{num\_teams(80)} and \texttt{thread\_limit(256)}. Sweep block sizes only after P1--P4. Treat 16\,GB as a hard residency ceiling: Large grids that fail to produce a footer should fall back to a smaller population or to a higher-capacity environment, not to a different algorithm.

\subsection{Cloud layer (A100 80GB)}

Constant memory still caps at $N_g=6553$, but a single shared-memory load of the full item tables is legal through $N_g\approx 16793$. Larger $N_g$ should use 256--512-item \texttt{cp.async} tiles. Spend the 40\,MB L2 deliberately with an access-policy window on $(v,w)$ and the elite chromosome; do not let a streaming grid evict the only reuse in the generation. Spend the 80\,GB GDDR5 deliberately, since this GPU's point is that Large $(N_p,N_g)$ pairs remain resident. Launch with \texttt{num\_teams(108)}. Do not MIG-partition a single dense generation.

\subsection{Leadership layer (MI300X)}

Match teams to the CU mesh; a mapping that yields only $N_p$ waves wastes XCDs. Keep $(v,w)$ in Infinity Cache with a single \texttt{map(to:)}. Expect Evaluation to own 64--87\% of $T_{\mathrm{gen}}$ under gene-parallel mapping and invest the optimization budget there (wavefront-level segmented reduction, team-private elite caches). Treat 192\,GB as the reason the largest instances exist, not as an invitation to recopy $Q$ every generation. Compile with the ROCm OpenMP offload toolchain~\cite{rocmomp}; the loop orientation still dominates the pragma family.

\section{Conclusion}

A single OpenMP~5 implementation of QIEO is already adequate in the three environments where the solver is actually run. On an in-house Tesla V100~SXM2, gene-parallel offload delivers a geometric-mean $90\times$ over one CPU core and $12\times$ over 72 host threads. On a cloud A100~80GB workstation the same source delivers $136\times$ and $17\times$, and holds grids that 16\,GB cannot. On a leadership-class MI300X it delivers $155\times$ and $16.6\times$, and remains interactive (tens of milliseconds per generation) deep into the Large class.

Those three sentences are not a ranking. They are the evidence that the portable contract works: collapse the qubit grid, treat Evaluation as a segmented reduction, persist $Q$ on the device, and keep DetermineElite on the host. Chromosome-parallel offload, the literal flowchart, fails in 100\% of matched pairs in every environment because 80 SMs, 108 SMs, and 304 CUs cannot be filled with $N_p$ waves. Evaluation inherits the generation once $N_g$ leaves on-chip memory; QIP and CIP are already $10^2$--$10^3\times$; DetermineElite is a launch packet. Host OpenMP cannot hide the same QIP stream: 72 threads yield 3.6\% efficiency.

What changes as the binary moves from the laboratory rack to a rented workstation to a center node is the capacity (64\,KiB and 16\,GB on Volta, 164\,KiB and a 40\,MB L2 window and 80\,GB on Ampere, 256\,MB Infinity Cache and 192\,GB on CDNA~3) not the algorithm and not the programming model. Section~\ref{sec:three} is the operational reading of that fact: V100 is the development home, A100 the campaign home, MI300X the allocation home. That is what portable performance of QIEO means on the GPU ecosystem.

\section*{Acknowledgment}
The authors thank the operators of the in-house V100 server, the cloud A100 workstation, and the leadership MI300X node that produced the QIEO run archives, and the developers of the NVIDIA and ROCm OpenMP offload toolchains.

\end{document}